\documentclass[aps,prl,twocolumn,superscriptaddress]{revtex4}
\usepackage{epsfig}

\begin{document}


\title{
Noise-Immune Conjugate Large-Area Atom Interferometers}
\author{Sven Herrmann}
\author{Sheng-wey Chiow}
\affiliation{Physics Department, Stanford University, 382 Via
Pueblo Mall, Stanford, California 94305, USA}
\author{Steven Chu}
\author{Holger M\"uller}
\email{hm@berkeley.edu;
http://physics.berkeley.edu/research/mueller/}
\affiliation{Physics Department, Stanford University, 382 Via
Pueblo Mall, Stanford, California 94305, USA}
\affiliation{Department of Physics, 366 Le Conte Hall, University
of California, Berkeley, CA 94720-7300} \affiliation{Lawrence
Berkeley National Laboratory, One Cyclotron Road, Berkeley, CA
94720.}

\date{\today}

\begin{abstract}
We present a pair of simultaneous conjugate Ramsey-Bord\'{e} atom
interferometers (SCI) using large (20$\hbar k$)-momentum transfer
(LMT) beam splitters, where $\hbar k$ is the photon momentum.
Simultaneous operation allows for common-mode rejection of
vibrational noise. This allows us to surpass the enclosed
space-time area of previous interferometers with a splitting of
20$\hbar k$ by a factor of 2,500. Among applications, we
demonstrate a 3.4\,ppb resolution in the fine structure constant
and discuss tests of fundamental laws of physics.
\end{abstract}


\maketitle 

\begin{figure}\centering
\epsfig{file=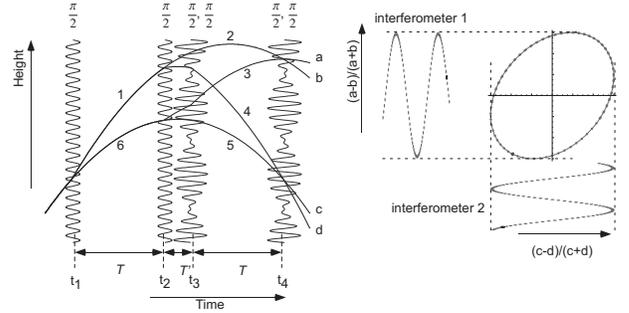,width=0.45\textwidth}
\caption{\label{schem} Correlating the fringes of two
interferometers creates an ellipse whose eccentricity allows to
determine the relative phase.}
\end{figure}

Light-pulse atom interferometers can convert a small signal into a
relatively large phase shift of the interference fringes. For
example, in Ref. \cite{LVGrav}, a 3 parts per billion (ppb)
modulation in local gravity leads to a 1\% shift of the
interference fringe. They thus make excellent microscopes for
small signals that have been applied in many cutting-edge
precision measurements
\cite{Peters,Gustavson,Durfee,Canuel,Fixler,Lamporesi,Wicht,Biraben,Dimopoulos,LVGrav}.
Large-momentum transfer (LMT) beam splitters, which have become
practical recently \cite{BraggInterferometry}, promise to increase
the sensitivity further, by factors of 10s to 100s, by increasing
the space-time area enclosed between the interferometer arms. But
just as vibrations blur microscopic images, they blur the
interference fringes in interferometers. This becomes more
pronounced as the sensitivity is increased until eventually
interferences can no longer be discerned. This has so far limited
the use of LMT beam splitters to very short pulse separation times
$T$ (Fig. \ref{schem}) of 1\,ms, thwarting the potential gain in
sensitivity.

Cancellation of vibrations has been demonstrated by using the same
laser light to simultaneously address two similar interferometers
at separate locations \cite{Snadden98,Stockton,Tino}. This method,
however, is restricted to situations where the differential signal
is small (in this case, the gravity gradient), so that the
interferometers are similar enough to be addressable by the same
laser.

In this work, we explore the full potential of LMT by cancelling
vibrations between dissimilar interferometers, the conjugate
Ramsey-Bord\'e interferometers shown in Fig. \ref{schem}. The idea
is to use laser pulses that contain a pair of frequencies whose
phase noise is extremely well correlated. We use this to
demonstrate a 2,500-fold increase in the enclosed space-time area
of interferometers with 20-photon momentum transfer, without a
reduction in contrast. This paves the path towards strongly
enhanced sensitivity in measurements of fundamental constants
\cite{Wicht,Biraben,Paris}, tests of general relativity
\cite{LVGrav} or the equivalence principle \cite{Dimopoulos}, and
detection of gravitational waves \cite{GravWav}.\\

\begin{figure}\centering
\epsfig{file=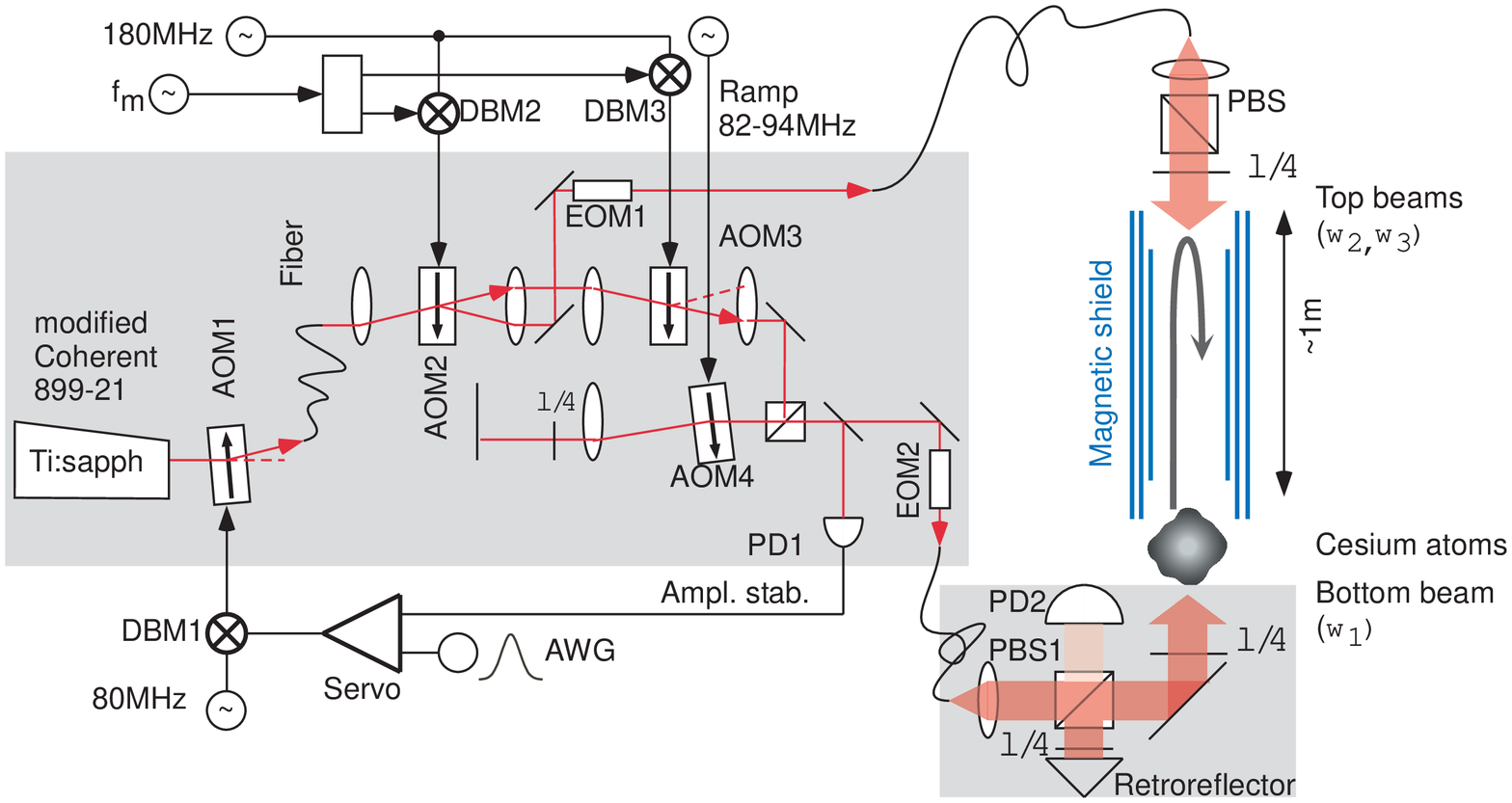,width=.45\textwidth}
\caption{\label{Setup} Setup.}
\end{figure}

Atom interferometers basically consist of a source of atoms and
beam splitters for the matter waves. The atom source for our
interferometers is a fountain of cesium atoms with a moving
optical molasses launch and Raman sideband cooling in an optical
lattice, as described in \cite{Treutlein}. As beam splitters, we
use multiphoton Bragg diffraction of matter waves at an optical
lattice \cite{Giltner,Miffre,Losses,BraggInterferometry}. The
optical lattice is formed by two counterpropagating laser beams
that we may call the top and bottom beam (Fig. \ref{Setup}). Bragg
diffraction can be described in the initial rest frame of the
atom. For example, the atom may absorb $n$ photons at $\omega_1$
from the bottom beam and be stimulated to emit $n$ at $\omega_2$
into the top beam. The atom emerges at the same internal quantum
state with a momentum of $2n\hbar k$, where $k$ is the wavenumber,
and a kinetic energy of $(n\hbar k)^2/(2M)$, where $M$ is the mass
of the atom. This energy has to match the energy
$n\hbar(\omega_1-\omega_2)$ lost by the laser field, which allows
us to choose the Bragg diffraction order $n$ by the difference
frequency $\omega_1-\omega_2$.

Fig. \ref{schem} (left) shows a space-time diagram of our
Ramsey-Bord\'{e} interferometers. Let us specialize to the lower
one, whose outputs are labelled c and d. An atom enters on its way
upwards. At a time $t_1$, a ``$\pi/2$" laser pulse transfers a
momentum of $2n\hbar k$ with a probability of 50\%. Depending on
whether momentum was transferred or not, the atom follows
trajectory 1 or 6. At $t_1+T$, a second $\pi/2$ pulse stops the
relative motion of them. After two more pulses, the paths are
recombined into the outputs c and d where they interfere. A
second, upper, interferometer is formed by recombining the other
outputs of the beam splitter at $t_2$.

The probability that the atom arrives at output c, for example, is
given by $\cos^2 \phi$, where $\phi=\phi_F+\phi_I$ is the phase
difference of the interferometer arms when they interfere. This
contains a contribution $\phi_F$ of the atom's free evolution
between the beam splitters and one of the interaction with the
light $\phi_I$. The free evolution phase $\phi_F=S_{\rm Cl}/\hbar$
is given by the classical action $S_{\rm Cl}=\int (E_{\rm
kin}-E_{\rm pot})dt$, where $E_{\rm kin}$ and $E_{\rm pot}$ are
the kinetic and potential energy. The interaction phase $\phi_I$
is because whenever a photon is absorbed, its phase is added to
the matter wave phase and subtracted for emission of a photon
\cite{Weiss}. This phase is different for the two paths because of
the respective spatial separation of the interactions at $t_2$ and
$t_3$. Summing up, \cite{Wicht,BraggInterferometry}
\begin{equation} \phi^\pm =\pm 8n^2\omega_r T+2n k
g(T+T')T+n\phi_L^\pm,
\end{equation}
where $\omega_r=\hbar k^2/(2M)$ is the recoil frequency and $g$
the local gravitational acceleration. The plus and minus signs are
for the upper and lower interferometer, respectively, and
$\phi_L^\pm=\phi_2-\phi_1-\phi_4^\pm+\phi_3^\pm$ is given by the
phases $\phi_{1-4}$ of the laser pulses at $t_{1-4}$. This
equation shows that LMT beam splitters can increase the
sensitivity of the phase towards gravity by a factor of $n$ and
the one towards the recoil by $n^2$.

Because of the motion of the atoms, which gives rise to a Doppler
frequency shift, addressing the upper and lower interferometer
requires two separate laser frequencies $\omega_2,\omega_3$ in the
top beam. The phases of these respective frequencies are denoted
$\phi_3^\pm$ and $\phi_4^\pm$.

The influence of vibrations is because the atoms define a freely
falling inertial frame. Any vibrations of the laboratory translate
into phase shifts of the laser beams in this frame, and if the
distribution of $\phi_L$ has a width that is comparable to
$\pi/n$, they render the interferences invisible. With non-LMT
beam splitters, vibrations can be suppressed to acceptable levels
by state of the art vibration isolation \cite{Hensley}. This
becomes difficult, however, with $n\gg 1$ LMT beam splitters.

The idea underlying the cancellation of vibrations is to run the
upper and lower interferometers simultaneously (``simultaneous
conjugate interferometers," SCIs) and to use
\begin{equation} \label{SCIphase} \Phi\equiv \phi^+-\phi^- =16n^2\omega_r
T+n\phi_L,
\end{equation}
where
\begin{equation} \phi_L=(\phi_3^+-\phi_3^-)+(\phi_4^--\phi_4^+)
\end{equation}
depends only upon the difference between laser phases at the last
two beam splitters. Thus, the requirement of {\em absolute} phase
stability in one individual interferometer has been reduced to one
of {\em relative} stability between two: if $\delta \phi_L=0$, the
overall phase $\Phi=$const., independent of vibrations. Then,
their fringes as plotted in Fig. \ref{schem} form an ellipse. The
common phase moves the data points around the ellipse, but the
differential phase can be extracted by ellipse-specific fitting.
One way that has been realized, without LMT
\cite{Snadden98,Stockton,Tino}, is to use the same radiation to
address both interferometers, which trivially leads to $\delta
\phi_L=0$. However, the use of this method is restricted to
measurements of very small differential signals, so that the
interferometers are similar; whereas our SCIs are dissimilar and
can be sensitive to the relatively large $16n^2\omega_r T$
differential signal.

One essential requirement for the experimental setup (Fig.
\ref{Setup}) is to satisfy $\delta \phi_L=0$ as well as
technically possible. Moreover, our laser system is optimized for
driving LMT beam splitters based on high-order Bragg diffraction
\cite{Losses,BraggInterferometry}, which requires laser pulses
having smooth envelope functions with an optimized duration and
high power.

The laser light originates from a 6\,W injection-locked
titanium:sapphire laser at 852\,nm wavelength
\cite{BraggInterferometry,6Wlaser}. Its frequency is referenced to
the $F=3\rightarrow F'=2$ D2 line of a modulation transfer
spectroscopy in a Cs vapor cell (not shown); an offset of up to
$\pm 20$\,GHz can be set by means of an offset lock. For intensity
control and forming the Gaussian envelope functions of the beam
splitting-pulses, we use an acousto-optical modulator AOM1 within
a feedback loop, see Fig. \ref{Setup}.

As a result of the different velocities, the resonance conditions
for the third and fourth beam splitter are shifted by
$16n\omega_r$ between the two interferometers. To satisfy both,
AOM2 is driven by two rf signals of equal amplitude at frequencies
of 180\,MHz$\pm f_m$. It thus generates two optical frequencies in
its deflected output that differ by $2f_m$. They follow the same
optical path; thus, length fluctuations such as caused by
vibrations, air currents, etc., are common-mode and do not degrade
the phase noise in the difference frequency. We have previously
shown that a phase variance of $\sigma^2 \approx
(160\,\mu$rad$)^2$ is possible \cite{PLL}. The power in each
component is set to 1/8 of the power at the input of AOM2, which
maximizes the Rabi frequency of driving the atoms.

To generate the counterpropagating beam, we use the undeflected
power from AOM2. Use of this radiation, which would otherwise be
lost, allows us to make good use of our available laser power.
This, however, varies between 1/2 and 1 of AOM2's input power at
the beat frequency $2f_m$ of the two rf signals. AOM3 is used to
take out this undesired modulation. It is driven by a ``conjugate"
rf signal, which is strong when the rf drive of
AOM2 is weak and vice versa. 
The amplitude modulation is thus suppressed - residual sidebands
are below 0.0018 (or -27\,dB) of the carrier power.

Due to the free fall of the atoms, the resonance condition in the
laboratory frame changes at a rate of about 23\,MHz/s. We account
for this by ramping the frequency of the bottom beam  using the
double-passed AOM4.

The beams are brought to the experiment via single-mode,
polarization-maintaining fibers and collimated to an $1/e^2$
intensity radius of about 3\,mm by a commercial fiber port
(Thorlabs, Inc., Newton, NJ) or 12.5\,mm by a triplet lens, a
combination of an aplanatic meniscus (CVI Melles Griot 01 LAM
225/076) and an achromatic doublet (Thorlabs AC508-200B). The
polarizations are made circular to ($\sigma^+ - \sigma^+$) by
zeroth-order quarter wave retardation plates. The bottom beam can
have a maximum power of 1.15\,W at the fiber output; the top beam
a peak power of 1.6\,W, i.e., 0.4\,W per frequency. Alternatively,
we overlap both beams at a polarizing beam splitter and send them
through the same fiber with orthogonal polarizations. The upper
fiber collimation optics is then replaced by a hollow corner-cube
retroreflector and a quarter wave plate. This method simplifies
beam alignment. Also, it was found important to shield the beams
from air currents to prevent a loss of contrast.

The first interferometer pulse is typically applied at
$t_1=70$\,ms after launch. For $20\hbar k$ beam splitters, we use
a detuning of about 3-4\,GHz and peak intensities of $
0.4\,$W/cm$^2$ in the bottom beam and $0.13\,$W/cm$^2$ per
frequency in the top beam with a waist of 12.5\,mm; with thin
($w_0=3$\,mm) beams a detuning of 16\,GHz is used. After elapse of
the full interferometer sequence, the atoms in the four
interferometer outputs a-d (Fig. \ref{Setup}, left) are separately
detected by their fluorescence $f_{a-d}$ as they pass a
photomultiplier tube in free fall. To take out fluctuations in the
atom number, we define the normalized fluorescence
$F_u=(f_a-f_b)/(f_a+f_b)$ of the upper interferometer and $F_l$ in
analogy for the lower interferometer.

\begin{figure}\epsfig{file=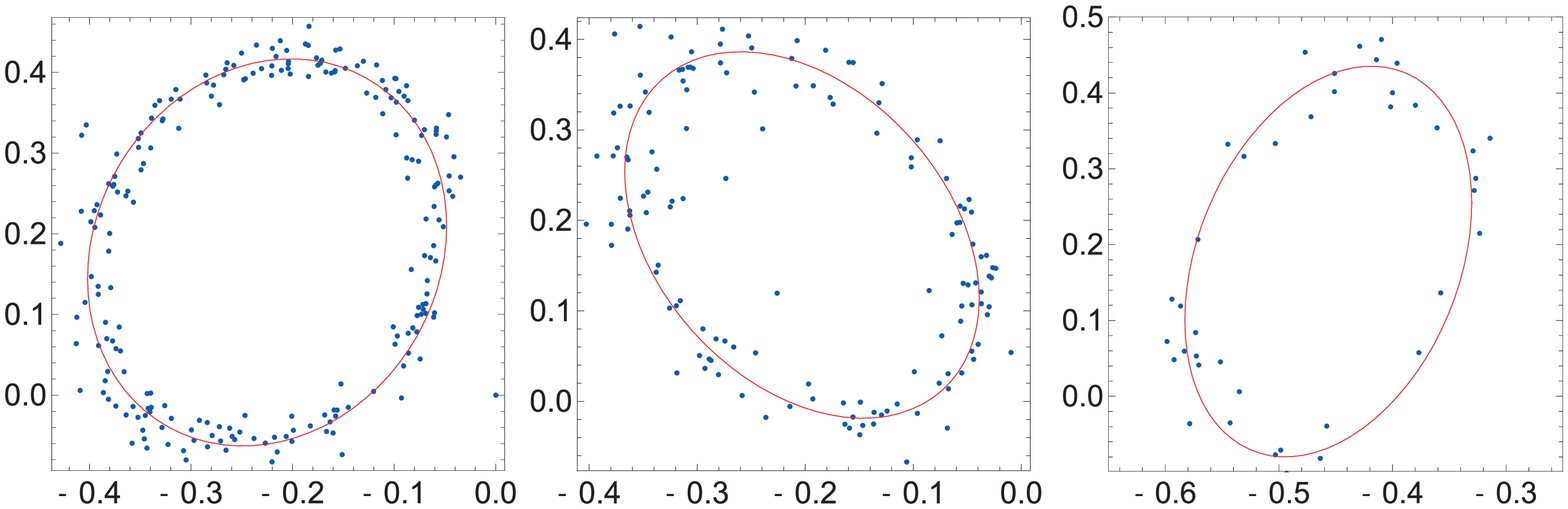,width=0.45\textwidth}
\caption{\label{Ellipses} Left: 12$\hbar k$, 1ms, 25\% contrast.
Middle: 14$\hbar k$, 1ms, 25\% contrast; Right: 20$\hbar k$, 1ms,
27\%.}
\end{figure}

Fig. \ref{Ellipses} shows examples for ellipses measured by our
SCIs at a short pulse separation time of $T=1$\,ms. A contrast of
around 25-31\% is achieved at momentum transfers between
$(8-20)\hbar k$ (the theoretical optimum is 50\%, because each
detected interferometer output overlaps spatially with population
lost in the third beam splitter which does not interfere). It is
evident that the strong dependence of the contrast upon the
momentum transfer, that was observed in previous LMT
interferometers \cite{BraggInterferometry}, is absent. This is a
first benefit of SCIs.

\begin{figure}
\epsfig{file=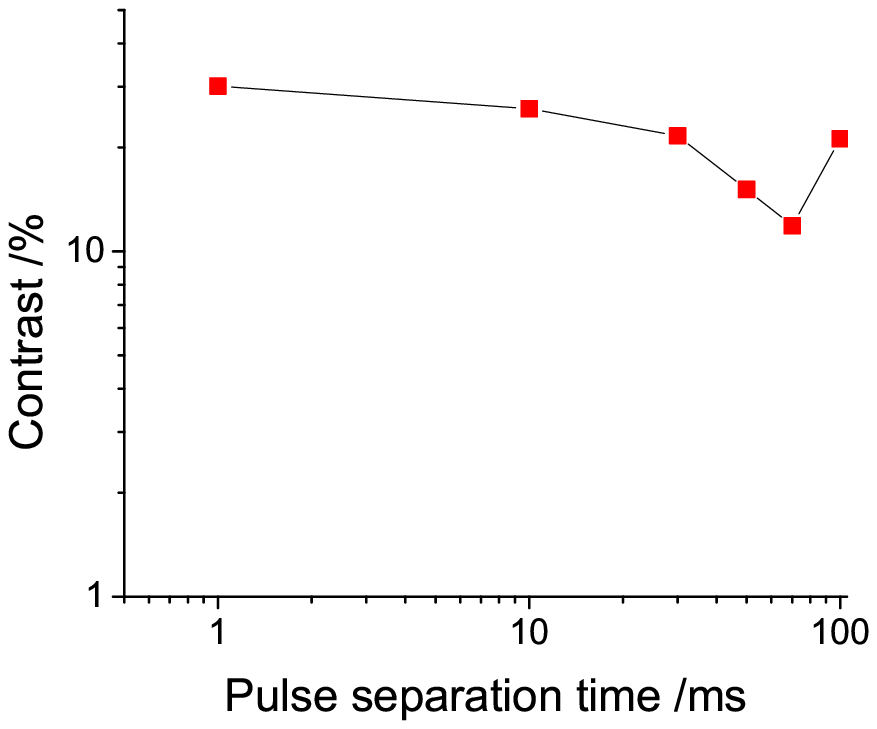,width=0.23\textwidth}
\epsfig{file=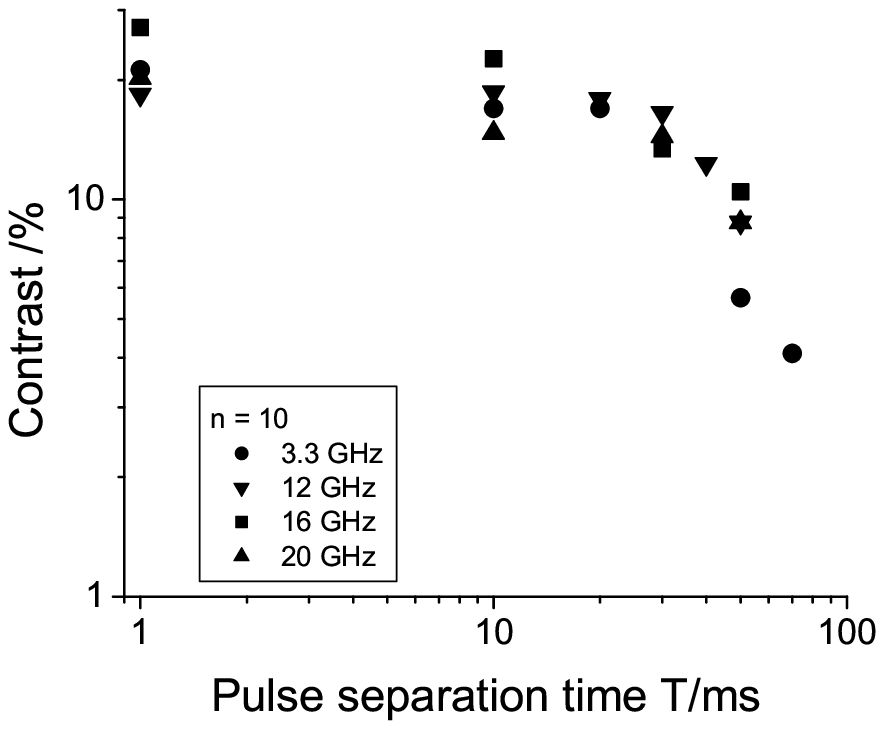,width=0.23\textwidth}
\caption{\label{10thin} Left: Contrast versus pulse separation
time for a $10\hbar k$ interferometer. The improvement at
$T=100\,$ms was reached by improved optics and shielding from air
currents (see Fig. \ref{Data} for the ellipse underlying this
datum). Right: same for a $20\hbar k$ interferometer with
different detunings.}
\end{figure}

The dependence of the contrast on the pulse separation time $T$ is
shown for $10\hbar k$ and $20\hbar k$ interferometers in Fig.
\ref{10thin}. A certain decrease for long $T$ is expected because
then a fraction of the atoms leave the area of the laser beams due
to their thermal velocity. Nevertheless, a contrast of $21\%$ can
be obtained for $T=100\,$ms and $10\hbar k$. For a $20\hbar k$
device, contrast is $10\%$ at $T=50\,$ms. In previous work without
SCIs, 8\% contrast at 20$\hbar k$ was only possible at $T\leq
1$\,ms \cite{BraggInterferometry}. Thus, SCIs allow us to improve
the pulse separation time to 50\,ms from 1\,ms, without a
reduction in contrast. This corresponds to a 2,500-fold increase
in the enclosed space-time area.

\begin{figure}
\epsfig{file=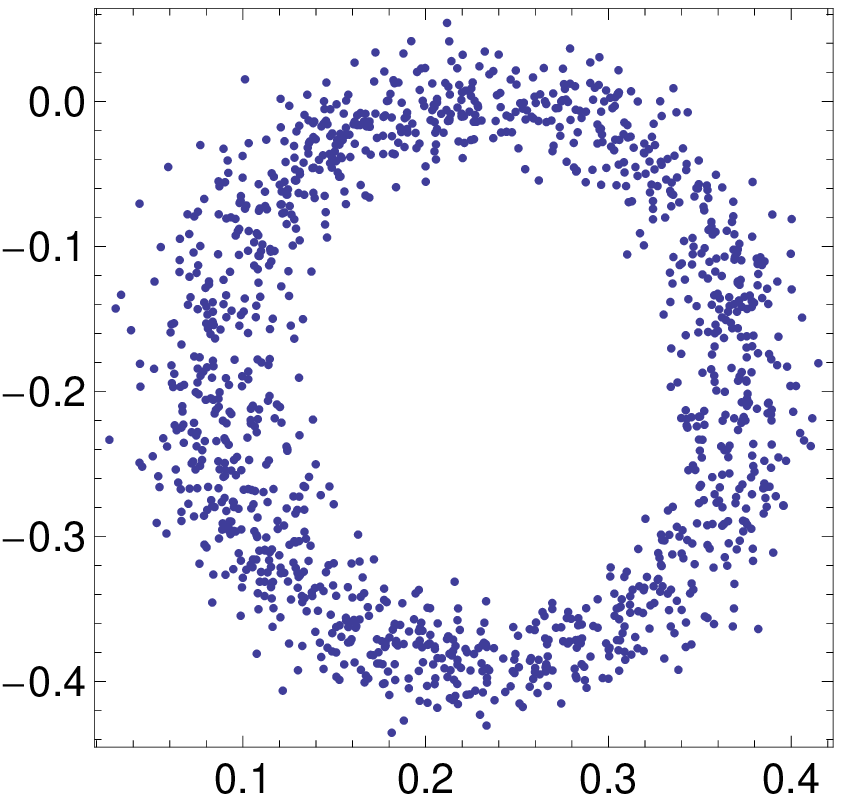,width=0.2\textwidth}
\epsfig{file=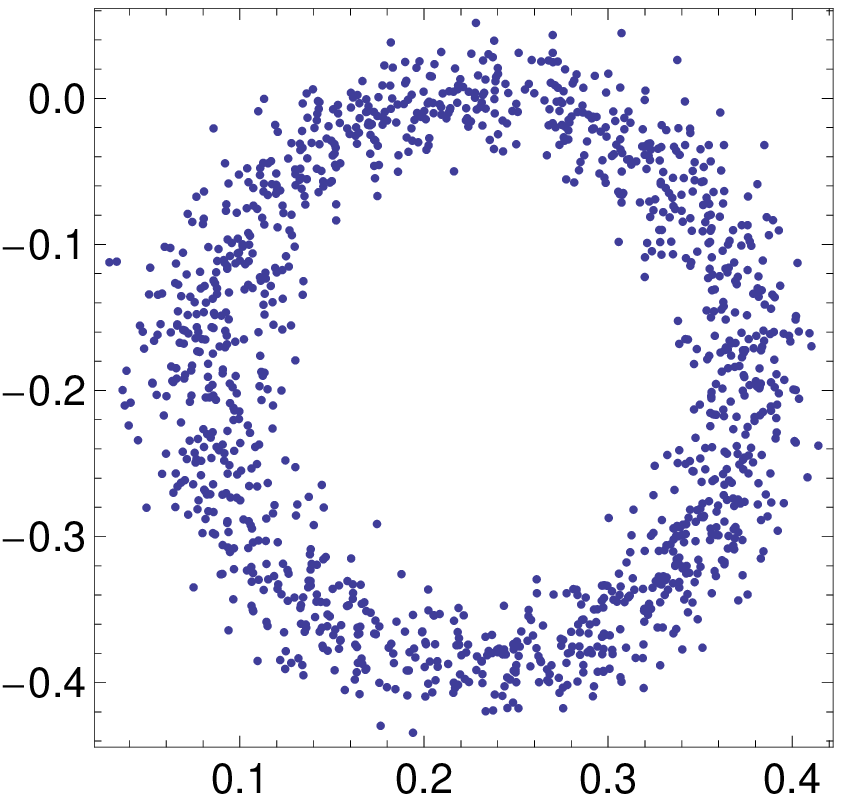,width=0.2\textwidth}
\caption{\label{Data} Each graph shows 1300 data points ($n=5,
T=100$\,ms, $C=20\%$) taken at $\Phi=+\pi/2$ and $-\pi/2$,
respectively.}
\end{figure}

Our dissimilar interferometers exhibit a large differential signal
(the $16n^2\omega_r T$ term due to the photon recoil), allowing
for high-resolution measurements. To demonstrate this, Fig.
\ref{Data} shows 2 data sets, each containing 1,300 data points
taken with a $10\hbar k$ interferometer pair having a pulse
separation time of $T=100\,$ms. To analyze the data, we use a
Bayesian estimation \cite{Stockton}, which shows a better immunity
from systematic errors than simpler methods \cite{ellipfit}. At a
given signal to noise ratio, the phase $\Phi$ can be best
determined if it is near $\pm \pi/2$, when the ellipse is close to
a circle. Therefore, an offset of $+\pi/2$ and $-\pi/2$,
respectively, was used for the two measurements shown, and the
laser frequency offset $f_m$ that yields $\Phi=0$ can be
determined from the average of the phase estimates. From a total
of 12,000 such points, we obtain a resolution of 6.8\,ppb within
7\,h of measurement. Via Eq. \ref{SCIphase}, $\omega_r$ and thus
$\hbar/M$ can be determined; correspondingly, our SCIs are
sensitive to the fine structure constant $\alpha$ via
$\alpha=[(2R_\infty/c)(M/m_e)(h/M)]^{1/2}$, where $R_\infty$ is
the Rydberg constant and $m_e$ the electron mass, to a resolution
of 3.4\,ppb. Without SCIs and LMT, achieving a similar resolution
would take several weeks' worth of data \cite{Wicht}. It is
similar to interferometers using Bloch oscillation for $\sim 2000
\hbar k$ common-mode momentum transfer, where 88\,h of data
($4\times 221$ fringes that take 6 min each) yield 3\,ppb
resolution and 4.6\,ppb absolute precision \cite{Biraben}. This is
because of (i) the $n^2$ scaling of the sensitivity with momentum
transfer in our method, and (ii) cancellation of vibrational noise
between SCIs. Further reduction of this statistical uncertainty
and an analysis of systematic errors are beyond the scope of this
Letter.

We remark that cancellation of vibrations is equally important in
Mach-Zehnder interferometers (MZIs) with LMT. The adaption of our
methods to this case is straightforward, as MZIs are slightly
simpler, requiring three light pulses instead of four and
featuring a 100\% theoretical contrast. Many MZI applications gain
sensitivity proportional to the enclosed area, which means that
our work allows for a 2,500 fold improvement there. The vibration
cancellation between LMT interferometers demonstrated here is also
a crucial technology for the detection of gravitational waves
\cite{GravWav}.

The dissimilarity of the interferometers in this case could be
external fields affecting the interferometer geometry, different
atomic species, or even different laser wavenumbers $k_1$ and
$k_2$. In the latter case, the phase relationship between these
lasers could be established by a frequency comb. Cancellation of
vibrations then requires that $k_1n_1$ and $k_2n_2$, where
$n_{1,2}$ are the Bragg diffraction orders, satisfy a simple
rational relationship. A Lissajous figure will then be generated
which reduces to an ellipse for $k_1n_1=k_2n_2$. Bayesian
estimation can be used to extract the phase. The possibility of
correlating signals from different atoms are interesting for tests
of the equivalence principle \cite{Dimopoulos}, and may allow new
paths to cancel systematic effects in searches for an electron
electric dipole moment \cite{ChengEDM}, tests of charge neutrality
\cite{Neutr}, and other experiments.

In this work, we have presented common-mode rejection between
dissimilar atom interferometers addressed by different overlapped
laser frequencies. Compared to previous work
\cite{BraggInterferometry}, we demonstrate a 2,500-fold increase
in the enclosed space-time area of atom interferometers using
$20\,\hbar k$ momentum transfer, without a reduction in
interference contrast. By removing the most important limitation
on the space-time area, and hence sensitivity, of such large
momentum transfer interferometers, this work opens the door
towards many exciting experiments.

S.H. and H.M. thank the Alexander von Humboldt foundation. This
material is based upon work supported by the National Science
Foundation under Grant No. 0400866 and by the Air Force Office of
Scientific Research under Award Number FA9550-04-1-0040.

\end{document}